# All-optical control of ferromagnetic thin films and nanostructures

C-H. Lambert[1,2], S. Mangin[1,2], B. S. D. Ch. S. Varaprasad[3], Y.K. Takahashi[3], M. Hehn[2], M. Cinchetti[4], G. Malinowski[2,], K. Hono[3], Y. Fainman[5], M. Aeschlimann[4], and E.E. Fullerton[1,5]

1) Center for Magnetic Recording Research, University of California San Diego, La Jolla, CA 92093-0401, USA

2) Institut Jean Lamour, UMR CNRS 7198 –Université de Lorraine- BP 70239, F-54506 Vandoeuvre, France

3) Magnetic Materials Unit, National Institute for Materials Science, Tsukuba 305-0047, Japan

4) Department of Physics and Research Center OPTIMAS, University of Kaiserslautern, Erwin Schroedinger Str. 46, 67663 Kaiserslautern, Germany

5) Department of Electrical and Computer Engineering, University of California San Diego, La Jolla, CA 92093-0401, USA

The interplay of light and magnetism has been a topic of interest since the original observations of Faraday and Kerr where magnetic materials affect the light polarization. While these effects have historically been exploited to use light as a probe of magnetic materials there is increasing research on using polarized light to alter or manipulate magnetism. For instance deterministic magnetic switching without any applied magnetic fields using laser pulses of the circular polarized light has been observed for specific ferrimagnetic materials. Here we demonstrate, for the first time, optical control of ferromagnetic materials ranging from magnetic thin films to multilayers and even granular films being explored for ultra-high-density magnetic recording. Our finding shows that optical control of magnetic materials is a much more general phenomenon than previously assumed. These results challenge the current theoretical understanding and will have a major impact on data memory and storage industries via the integration of optical control of ferromagnetic bits.

The dynamic response of magnetic order to ultrafast external excitation is one of the more fascinating issues of modern magnetism [1, 2]. Optical probing at the femto-second time scale allows investigating ultrafast magnetization dynamics including different fundamental interactions between spins, electrons, and lattice degrees of freedom when materials are far from equilibrium [1, 3-5]. It further has the opportunity to explore the potential for novel technologies such as heat-assisted magnetic recording (HAMR) [6, 7]. An interesting and important outcome from studies of ultra-fast dynamics of magnetic systems is the demonstration that circularly polarized light can directly switch magnetic domains without any applied magnetic field. In their pioneering work the Rasing group in Nijmegen showed fully deterministic magnetization switching in a ferrimagnetic GdFeCo alloy film using circularly polarized femtosecond laser pulses [8]. This phenomenon is now referred to as All-Optical Helicity-Dependent Switching (AO-HDS).

Since this initial experimental discovery, there have been extensive explorations of these phenomena with a particular attention on the ferrimagnetic nature of the magnetic materials that, until now, have been the only materials to show AO-HDS. While initial studies were focused on rare-earth transition-metal (RE-TM) GdFeCo alloys there have been recent examples of AO-HDS in other RE-TM materials such as TbCo [9], TbFe [10], DyCo and HoCo alloys, Tb/Co and Ho/Co multilayers as well as Co/Ir/CoPtNiCo/Ir synthetic ferrimagnets [11]. In all these cases, AO-HDS was observed for ferrimagnetic systems with two distinct sublattices that are antiferromagnetically-coupled and exhibit a compensation temperature near or above room temperature. The initially discussed models for AO-HDS were based on the existence of an effective field created by the circular polarized light via the inverse Faraday effect [12, 13] or directly by the transfer of angular momentum from the light to the magnetic system [14]. More recent models are focused on the formation of a transient ferromagnetic state due to different demagnetization times for RE and TM sub-lattices and the transfer of angular momentum both between the different magnetic sub-lattices and the lattice. In these latter models the light's helicity plays a secondary role, via the helicity dependence of the absorption (*i.e.* circular dichroism) and no transfer of angular momentum and any coherent effects from the light to matter is required [15-17]. These models have been supported by recent measurements in GdFeCo alloys [18]. Further, there is also emerging evidence that laser-induced superdiffusive spin currents can flow in heterogeneous systems [17, 19-22]. It is suggested that angular momentum is removed during demagnetization by a flow of spin-polarized currents leading to a transfer of angular momentum between different lateral regions of the sample potentially contributing to the AO-HDS process [20].

Besides the still unclear microscopic explanation of AO-HDS the most intriguing and important open question is whether AO-HDS is specific to a subset of ferrimagnetic materials or is a fundamentally general process and can be applied to much more widely used ferromagnetic materials. Furthermore can it be also applied to reverse technologically important high-anisotropy granular or patterned materials that are anticipated for future high-density magnetic recording [23]? In this paper we demonstrate first that AO-HDS occurs for a range of ferromagnetic thin films with perpendicular magnetic anisotropy including Pt/Co/Pt trilayers and Co/Pt, Co/Pd, $Co_{1-x}Ni_x$/Pd and Co/Ni multilayers. In these cases we only observe AO-HDS for films where the magnetic film thickness is less than ~3 nm and this thickness appears limited by the demagnetization energies that drive domain formation during heating by the laser pulses. We then show a high degree of optical control of 15-nm thick granular FePt films currently being pursued for HAMR media which exhibit a room-temperature coercive fields exceeding 3.5 T [24, 25]. The level of control in the granular FePt case is determined by thermal activation of the grains after the application of the optical pulse.

To probe the optical response of ferromagnetic samples we use an optical/heat-assisted magnetic switching facility with a 100-femtosecond pulsed laser source (see Methods section and supplementary information for details). Shown in Figs. 1a-c are Faraday microscope images of laser line scans for [Co(0.4 nm)/Pt(0.7)]$_N$ multilayers where $N$=8, 5 and 3 for 1a, 1b and 1c, respectively. The laser is scanned across the sample and final magnetic configuration is subsequently imaged. For each figure the laser helicity is either right circular polarization (σ+), left circular polarization (σ-) or linear polarization (L) as labeled in the image. The samples have perpendicular magnetic anisotropy so the magnetization easy axis is normal to the film surface and the contrast results from the two possible directions of the magnetization. For Figs. 1a-c, the left hand of the image is magnetized up while the right is magnetized down with a domain wall that runs vertically in the middle of each image.

For $N$=8 (Fig. 1a) we observe domain formation where the region scanned by the laser is filled with stripe subdomains that minimizes the dipole energy [26]. This process is independent of the light polarization and we describe it as laser-induced thermal demagnetization (TD). A rim is observed at the edge of the scanned area where the magnetic orientation is opposite to the background and is stabilized by the dipolar fields arising from the surrounding film that supports the opposite direction of magnetization. For $N$=5 (Fig. 1b), we again observe the formation of subdomains in the scanned region. However, the average domain size is much larger than in Fig. 1a. This increase in domain size is expected for decreasing number of layers since the equilibrium domain size increases exponential with decreasing film thickness in the thin-film limit (see Ref. [26] and references within). More importantly we observe that the resulting domain structure depends on the light polarization. For σ+ light we observe white isolated bubble-like domains in a dark background while for σ- we observe isolated dark domains in a white background. For linear polarization we observe symmetric domain formation. We further see that the magnetization near the edges of the line scan again tends to favor the direction opposite of the surrounding film similar to that observed for $N$=8.

For $N$=3 (Fig. 1c) we observe something intrinsically different. We observe fully deterministic magnetization reversal of the material under the beam with no external magnetic field for both left and right circular polarization. In this case the orientation of the magnetization after the laser has passed depends solely on the helicity of the laser. This is the clear demonstration of AO-HDS in a ferromagnetic material. The process is reversible with reversing the helicity of the light and the final magnetization orientation can be related to the light helicity. Surprisingly the light absorption in the Co/Pt multilayer samples that show AO-HDS is lower when the light has the circular polarization needed for switching (*i.e.* the magnetization switches into the high absorption state). This observation is opposite to the dichroism-induced switching discussed in Ref [27]. Finally, the domains created in the case of linear polarization are much bigger for $N$=3 in accord with the small dipolar energy gain by domain formation in this case [26].

Shown in Fig. 1d are images of domain patterns for the $N$=3 sample for various laser powers where the film is saturated in one direction and the laser spot is fixed and not scanned on the surface. We see that for low power (362 nW) a reversed domain is written for right circular polarization while there is no change to the film for left circular polarization. A region of random domains is observed for linear polarization. With increasing laser power from right to left, regions of demagnetized random domains develop in the center of the laser spot for all three polarizations indicating that the power is such that the temperature exceeds a critical temperature for which domains are formed. This can result from exceeding the Curie temperature ($T_C$) or from a critical temperature where there is a loss of perpendicular anisotropy or enhanced domain fluctuations. However for right circular polarized light there is a rim at the edge of the demagnetized area that shows deterministic switching that is not present for left circular or linear polarized light. The rim is not visible for the left circular polarization as it is in the same direction as background film. If the magnetization is reversed the rim for left circular polarized light is observed. This rim is similar to what has been previously observed for ferrimagnetic films (Fig. 2 of Ref [8]) and indicates a window of laser power for AO-HDS as the laser power decreases moving away from the center of the spot beam.

We further explored the effective driving energy for AO-HDS and domain formation by adding external magnetic fields to the experiments shown in Figs. 1a-c (see supplementary Figs. S1-S3). While linear polarization leads to domain formation in zero applied field the application of a magnetic field can stabilize a uniform magnetization state. This field increases from 3-4 Oe for the $N$=3 sample to ~12 Oe for $N$=5 and to ~40 Oe for $N$=8 demonstrating the increased demagnetization energy with thickness. In other word, for increasing magnetic thickness larger applied fields are needed to suppress domain formation after heating with the laser. When an applied field is combined with circular polarization the applied field can either support or oppose the circular polarization. For $N$=3 a field of 7 Oe is needed to oppose the circular polarized light and yield a demagnetized film while a field of ~12 Oe will saturate the film in the opposite direction as that expected for the helicity of the light (Supplementary Fig. S3). For $N$=5 the field to yield a demagnetized film is ~12 Oe while a field of ~25 Oe is needed to saturate the film opposite to the light helicity (Supplementary Fig. S2). However, when comparing the effects of circular polarization of the light to the applied magnetic field in these experiments one has to remember that the field is applied during the entire thermal process while the role of the helicity of the pulse persists only for a few picoseconds after the laser excitation [28].

We have explored a range of thin ferromagnetic film materials to determine how general the phenomena shown in Fig. 1c is by studying $[Co(t_{Co})/Pt(t_{Pt})]_N$, $[Co(t_{Co})/Pd(t_{Pt})]_N$, $[Co_xNi_{1-x}(0.6nm)/Pt(0.7nm)]_3$ and $[Co/Ni]_N$ multilayer structures where we have varied several material parameters (*e.g.* $t_{Co}$, $t_{Pt}$, $N$ and Ni concentration). In short we observe AO-HDS in all these ferromagnetic materials classes including single Co layers sandwich between two Pt layers. Shown in Fig. 2 are selected results for the threshold laser power needed to achieve either AO-HDS (solid symbols) or TD (open symbols). Figure 2a are results for $[Co(t_{Co})/Pt(0.7nm)]_N$ where the threshold laser power increases with both $N$ and $t_{Co}$. These results show AO-HDS for $N$=2 or 3 and $t_{Co} \leq 0.6$ nm (*i.e.* the thinnest samples) and TD for thicker samples (consistent with Fig. 1). The trends of the threshold powers are independent of AO-HDS or TD processes and increasing linearly with either $t_{Co}$ or $N$. This suggests that the two phenomena are linked by a common mechanism or similar temperatures are needed for both processes.

Figure 2b are the results for $[Co(0.4 nm)/Pt(t_{Pt})]_2$ samples where we tune the Pt thickness for fixed $N$ and $t_{Co}$ (we also measured $[Co(0.4nm)/Pd(t_{Pd})]_2$ samples as a function of Pd thickness and observed similar results). This increased the thickness of the film but leaves the total magnetic moment relatively unaffected. This also dramatically changes the exchange coupling between the Co layers since there is induced ferromagnetic moments in the interfacial Pt atoms. As can be seen in Fig. 2b the threshold power decreases slightly with increasing Pt thickness. For $t_{Pt} = 1.2$ nm (Fig. 2b) the two Co layer are only weakly coupled suggesting that single Co layers may also switch. Shown in Fig. 2c are the results for $N$=1 trilayer structures (*i.e.* a Pt/Co($t_{Co}$)/Pt structures) where we increase the Co layer thickness. We observed AO-HDS for samples 0.6 nm $\leq t_{Co} \leq$ 1.5 nm. The upper limit is set by the thickness where the sample maintains perpendicular magnetic anisotropy. The lower limit is set by the sensitivity of the optical detection. Again, the threshold power increases linearly and we observe AO-HDS for a single ferromagnetic film. In fact the threshold values for a single Co layers are consistent with the extrapolation the data in Fig. 2a to $N=1$. Figure 2d shows the threshold power for $[Co_{1-x}Ni_x(0.6nm)/Pt(0.7nm)]_N$ multilayers as a function of both $N$ and Ni concentration. We find the threshold power increases with N as seen in Fig. 2a and decreases with Ni concentration and the trends are independent of TD or AO-HDS. Finally we observe AO-HDS for these following Co/Ni structure Ta(3nm)/Cu(10nm)/[Ni(0.5nm)/Co(0.1nm)]$_2$/Ni(0.5nm)/Cu(5nm) where perpendicular anisotropy arises primarily from the Co-Ni interfaces.

These results show that AO-HDS is a rather general phenomena for magnetic films but seem to be limited to the thin-film limit. Such structures are useful in a number of spintronic applications (e.g. magnetic random access memory). However applications such as high-density magnetic recording require small magnetic grains or patterned bits for high signal-to-noise readback of the data and high anisotropy to remain thermally stable at the nanoscale [29]. The current challenge is that the magnetic fields required to write high-anisotropy grains is above what can be achieved by electromagnetic write heads. HAMR is the leading technology to address the challenge where a laser spot heats the magnetic material close to $T_C$ where the anisotropy field is lowered sufficiently to allow the grains to be written with an external magnetic field [6, 7]. Further using the polarization of the light to directly write the bits or to supplement the write field would greatly simplify the design of the write elements.

To explore this issue we have studied the role of AO-HDS on high-anisotropy granular FePt-based films being developed as a candidate media for high-density HAMR [24]. We studied both FePtAgC and FePtC granular media grown onto single-crystal MgO. The preparation method leads to the formation of high-anisotropy FePt grain separated by C grain boundaries. The average FePt grain size is ~9.7 nm and ~7.7 nm for the FePtAgC and FePtC granular media, respectively. The room-temperature perpendicular magnetic anisotropy and coercive fields are 7 T and 3.5 T, respectively, for both films. Plan-view electron microscope images and magnetization characterization are shown in supplementary Figs. S4 and S5. Shown in Fig. 3 are results of optical studies for the FePtAgC granular film where we start with the film in a random magnetic state with equal up or down oriented magnetic grains (similar results are obtained for the FePtC film). Because the grain size is well below the resolution of the Faraday microscope the randomly magnetized sample appears grey. As can be seen from Fig. 3 there is a clear net magnetization achieved that depends on the helicity of the circularly polarized light and no change is observed in the image with linear polarization. This clearly shows that a percentage of the films is being magnetized and controlled by the polarization of the light. Shown in Fig. 3b are images of the laser spot without scanning the laser beam similar to those in Fig. 1d. As shown there is a laser power to achieve AO-HDS which exists for a relatively narrow range of powers. With increasing power above the threshold power (~420 nW) there is a region of reversed grains. Above ~600 nW a ring forms where AO-HDS occurs at a particular radius (and this radius grows with increasing power). The center of the ring where the laser intensity is the highest the films is demagnetized, presumably from exceeding $T_C$.

While clear AO-HDS is observed the degree of magnetization is less than 100%. By comparing the contrast to the saturated film we can estimate that the induced magnetization is ~10-20 % of saturation. The lack of saturation can arise from at least two effects or the combination of these effects. The first is that AO-HDS is only affecting a subset of the grains. The second is that the AO-HDS is efficient and saturates the sample, but that the magnetic grains that make up the films sample are highly thermally activated and between the time of AO-HDS and the sample cooling the grain assembly partially demagnetizes due to thermal switching of the grains. When the sample is heated toward $T_C$ there is a strong drop in the magnetic anisotropy ($K_U$) near $T_C$. This is the basis of HAMR where this lowers the coercive field to a point where when a modest field is able to reverse the grains. However, at this point the energy stored in the grain $K_U V$ where V is the volume of the grain becomes comparable to thermal energy $k_B T$ and therefore there is a high probability for thermal reversal of the grains while the sample is cooling. This effect is further driven by the dipolar fields from the neighboring grains that support a demagnetized ground state. This process is described in detail in the literature, see for example Ref. [30].

To quantify the relative role of thermal activation we applied magnetic fields while the sample was illuminated by the polarized light. Shown in Fig. 4 are the results of line scans with both right ($\sigma+$) and left ($\sigma-$) circular polarization in increasing applied static magnetic field. The field direction is chosen so it supports the right circular polarized light and opposes the left circularly polarized light. We find that an applied field of ~700 Oe is sufficient to suppress the effects of the helicity of the light where no contrast is observed for left circular polarization in a 700 Oe field. For right circular polarization the contrast increases with increasing field. Similarly, we can excite the films with linear polarized light and an applied field of ~600 Oe is needed to obtain a similar magnetization as the AO-HDS results (see supplemental materials Fig. S6). The fact that these modest fields are sufficient to counter the polarization of the light indicates we are heating near $T_C$ where the small 700-Oe field (a factor of 50 less than the room temperature coercive field) can alter the magnetization orientation of the grains. Moreover the fact that applied fields up to 1100 Oe are not sufficient to fully saturate the film after the laser excitation indicates that the grains are highly thermally activated during optical excitation and we are observing stochastic processes. This is also consistent with measurements of the saturated film where the magnetization is decreased with any polarization, even for the circular polarization that supports the magnetization. Further control or deterministic switching may require careful engineering of the laser pulse shape and thermal response of the magnetic film and substrate through, for example, the introduction of heat sink layers.

Our results show that a ferrimagnetic structure is not necessary for AO-HDS to be observed and therefore antiferromagnetic exchange coupling between two magnetic sublattices is not required. However, these finding cannot rule out the role of two magnetic sublattices on the magnetic reversal since all of our examples have two magnetic elements that are ferromagnetically coupled. While the data shown in Fig. 2c is for a single Co film sandwiched by Pt, the Pt atoms at the interface are polarized by the Co and are magnetic and, therefore, these systems sample have two magnetic elements. This also applies for FePt films the Pt contributes to the magnetization. Given that we observe AO-HDS switching on single Co films as well as Co/Pt multilayers it is unlikely that super-diffusive currents that couple different magnetic regions in a heterogeneous sample is required. However, we cannot rule out that flow of currents into the Pt layer does not play a role.

Our results do suggest that we are heating near the Curie point and that this is important for the AO-HDS in ferromagnetic materials. The threshold intensities shown in Fig. 2 generally track with what is the expected trends for $T_C$ for these systems (increasing $t_{Co}$ or $N$ increases $T_C$ while increasing $t_{Pt}$ or Ni concentration decreases $T_C$). The final state is most likely determined by angular momentum transfer of the light to the magnetization or a resulting effective field from the light acting on the magnetization. This is expected to be most effective when approaching $T_C$ where even modest angular moment transfer, effective fields or applied magnetic fields can lead to a symmetry breaking such that magnetization is deterministically switched. This magnetization state will be maintained unless the demagnetization and thermal energies that favors domain formation are too large and cause demagnetization during cooling. AO-HDS is then expected to be generally observed as long as the energy gain by domain formation is sufficiently small or controlled to avoid demagnetization during cooling. For perpendicular magnetized films there are strong demagnetizing fields within the film that support domain formation. The energy gain for domain formation is strongly suppressed in the ultrathin film limit yielding increasing domain sizes with reduced thickness and explains the observation of AO-HDS only in the thin-film limit. A secondary way to avoid domain formation is using low magnetization materials. Note that such a description is consistent with previous measurements of ferrimagnetic alloys, multilayers and heterostructures where AO-HDS switching is generally observed when the compensation temperature (*i.e.* the temperature where the net ferrimagnetic moment is zero) is near or above room temperature. Having a compensation temperature between room temperature and $T_C$ will help suppress domain formation even for relatively thick films.

In conclusion we demonstrated the optical control of the magnetization of a variety of ferromagnetic materials (thin film, multilayers and granular media). These results demonstrate a new and technologically important class of materials showing AO-HDS phenomenon. By challenging the current theories in the field, this study offers significant progress toward a better understanding of the interaction between pulsed polarized light and magnetic materials. However it is clear that a number of questions still need to be addressed to gain a fundamental understanding of all the mechanisms involved. The control of the magnetic orientation of ferromagnetic thin film and granular media using light opens new applications in magnetic memory, data storage and processing. Given the current trends for silicon nanophotonics, miniaturization, and photonic-electronic integration, the ability to couple photonic, electronics, and magnetic materials will significantly extend the level of flexibility in existing devices and enabling completely new applications.

**Methods:**

We used optical pulses, having a central wavelength of 800 nm (1.55 eV), a pulse duration of about 100 fs at the sample position and a repetition rate of 0.1-1 kHz. A schematic of the optical test facilities where AO-HDS measurements were performed is shown in the Supplementary Fig. S7. The response of the magnetic film was studied using a static Faraday microscope with 1-μm resolution that to image the magnetic domains while or after the laser illuminates the sample. The helicity of the beam is controlled by a zero-order quarter wave-plate, which transforms linear polarized light (L) into circularly left- (σ+) or right-polarized light (σ-).The present measurements were performed at room temperature (RT) and the laser beam was swept at a constant rate of ~3-20 μm/s with the typical laser spot size of ~80 μm. The laser power was adjusted to achieve either TD or AO-HDS and varied from sample to sample. Typical laser powers for 1 kHz repetition rates are 0.05 - 2 μW. The threshold power scales with the repetition rate indicating it is the energy/pulse that determines the threshold power.

All the thin-film samples (with the exception of the FePt samples) were grown by DC magnetron sputtering from elemental sources onto room-temperature glass substrates coated with a thin Ta seed layer. Alloys were grown by co-sputtering where the source powers controlled the composition. Multilayers and heterostructures were formed by sequential deposition of layers. The samples were then covered with a thin Ta capping layer. The FePt-C and FePtAg-C granular films were fabricated by DC magnetron sputtering using Fe, Pt and C targets on MgO substrates [1]. The film stack was MgO (001) sub./[FePt-C(0.25)/FePt(0.15)]25/C(15). The number in the parenthesis is the thickness of the each layer and the unit is nm. Multilayers of FePt-C/FePt were deposited at elevated temperatures of 550ºC and C capping layer was deposited at RT. The volume fraction of C was about 28%. The composition ratio of Fe and Pt is 1:1. Silver contained samples also produced in similar procedure as mentioned above except ten atomic percent is added during the deposition. Due to high deposition rates of Ag, we added 10 seconds per minute. In this way, we controlled the atomic percentage. The film stack for silver containing film was MgO (001)/[FePtAg-C(0.3)/FePt(0.15)]22/C(15). The total thickness of FePtAg-C is about 10 nm. Magnetic moment and hysteresis measurements were performed using a vibrating sample magnetometry and magneto-optic Kerr effect measurements. Sample structures were characterized by x-ray reflectivity and transmission electron microscopy.

[1] B.S.D.Ch.S. Varaprasad, Y.K. Takahashi and K. Hono, Japanese patent application 2013-189727.

**Acknowledgements**

We would like to the Robert Tolley and Matthias Gottwald with help on sample fabrication and helpful discussions with Marco Menarini and Vitaliy Lomakin. This work was supported by the ANR, ANR-10-BLANC-1005 "Friends," and work at UCSD was supported by the Office of Naval Research (ONR) MURI program and a grant from the Advanced Storage Technology Consortium. It was also supported by The Partner University Fund "Novel Magnetic Materials for Spin Torque Physics" as well as the European Project (OP2M FP7-IOF-2011-298060) and the Region Lorraine. Work at the National Institute for Materials Science was supported by IDEMA-ASTC.

**Author Contributions** SM, MA, YS, and EEF designed and coordinated the project; C.-H. L. grew, characterized and optimized the thin films samples while B. S. D. Ch. S. V., Y. K. T., and K. H. developed and grew the FePt-based granular media. C.-H. L., S. M., Y. K. T. operated the Kerr microscope and the pump laser experiment. S. M. and E. E. F. coordinated work on the paper with contributions from C.-H. L, M. H., M. C., and G. M. with regular discussions with all authors.

**Figures**

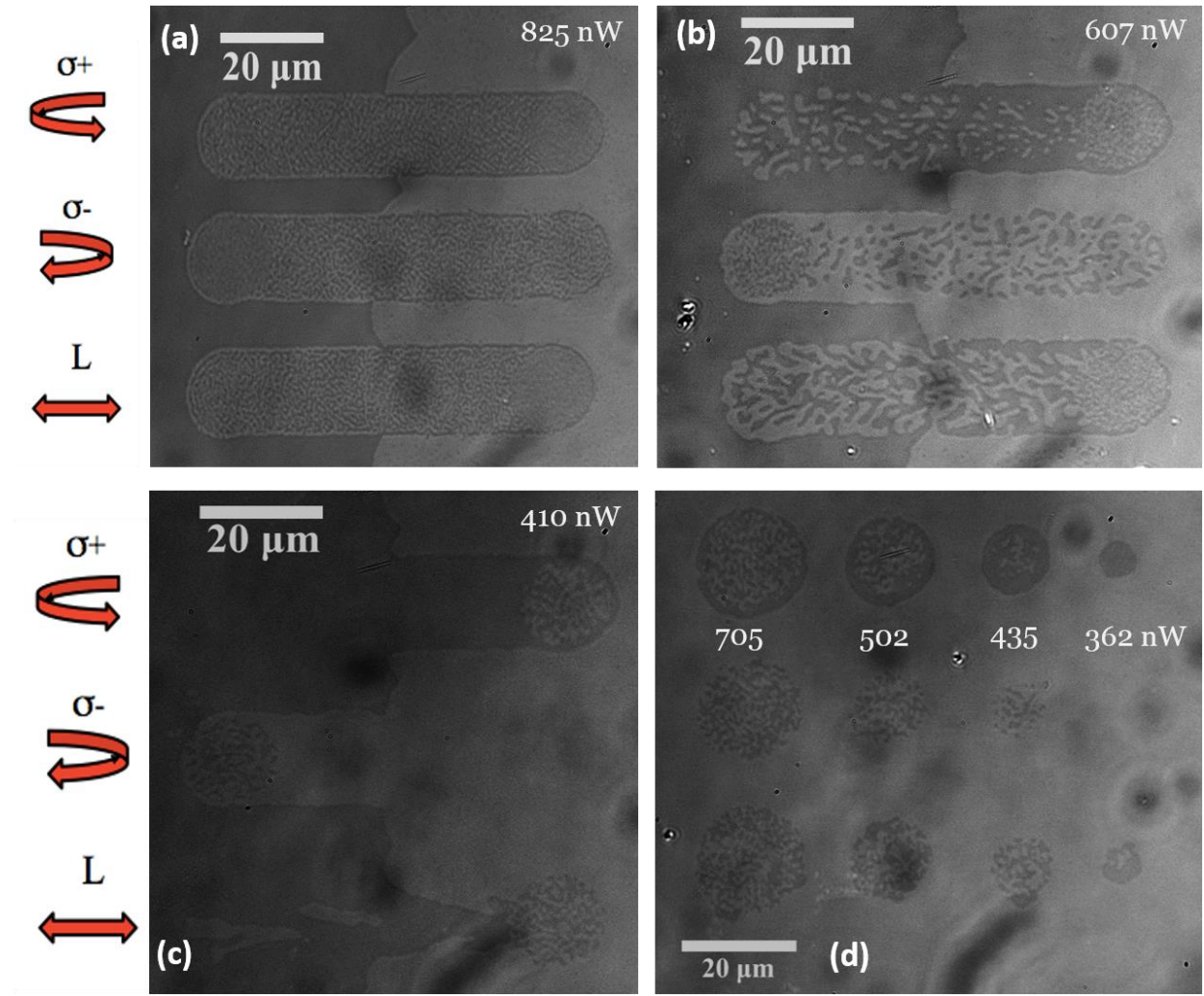

**Figure 1:** Magneto-photonic response of Co/Pt multilayers. Magneto-photonic response in zero applied magnetic field of [Co(0.4nm)/Pt(0.7nm)]$_N$ multilayer samples to various laser polarizations where (a) *N*=8, (b) *N*=8 and (c) and (d) *N*=3. For each image from top to bottom the laser polarization is right circularly polarized light (σ+), left circularly polarized light (σ-) or linear polarized light (L). For (a)-(c) the laser beam was swept over the sample with a magnetic domains and the magnetization pattern was subsequently imaged. In (d) the laser was fixed at individual spots where the average laser intensity increased as shown in the image.

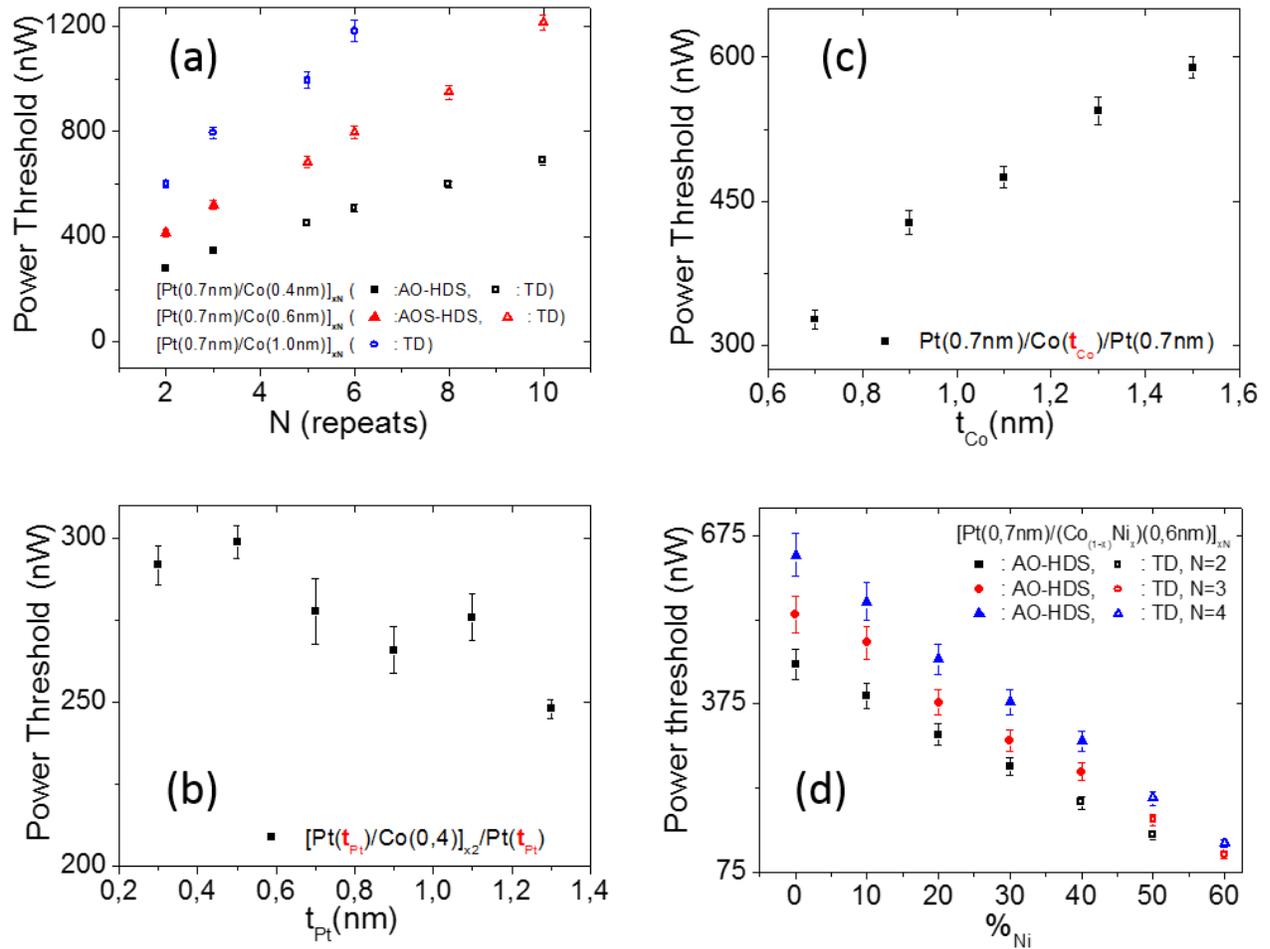

**Figure 2:** Magneto-photonic response of multilayer thin-film samples to circular polarization light and varying powers. Plotted is the evolution of the threshold power to achieve either thermal demagnetization (TD) or all-optical helicity-depenedent switching (AO-HDS) of various samples. For each sample the threshold power is given either as a filled symbol for AO-HDS or open symbols for TD. (a) Threshold power *vs.* N for [Co($t_{Co}$)/Pt(0.7 nm)]$_N$ samples with $t_{Co}$ = 0.4, 0.6 and 1.0 nm, (b) threshold power *vs.* $t_{Pt}$ for [Co(0.4nm)/Pt($t_{Pt}$)]$_2$ multilayer samples, (c) threshold power *vs.* $t_{Co}$ for Pt/Co($t_{Co}$)/Pt trilayer samples and (d) threshold power *vs.* Ni concentration for [Co$_{1-x}$Ni$_x$(0.6nm)/Pd(0.7nm)]$_N$ with N=2, 3 or 4.

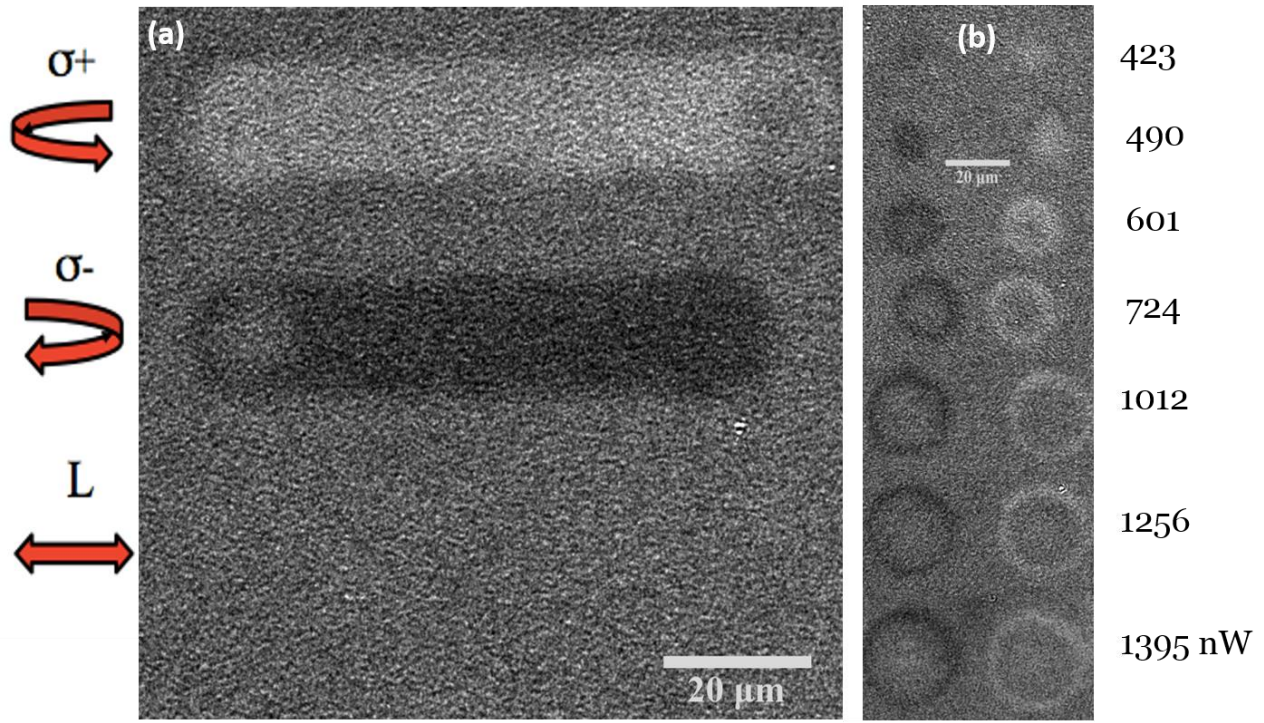

**Figure 3:** Magneto-photonic response in zero applied magnetic field of a 15-nm FePtAgC granular film sample starting with an initially demagnetized sample. (a) Line scans for various laser polarizations from top to bottom right circularly polarized light (σ+), left circularly polarized light (σ-) and linear polarized light (L). The laser beam was swept over the sample and the magnetization pattern was subsequently imaged. (b) Images of magnetic domains for right circularly polarized light (left column) and left circularly polarized light (right column) laser spot for various laser powers shown next to the image.

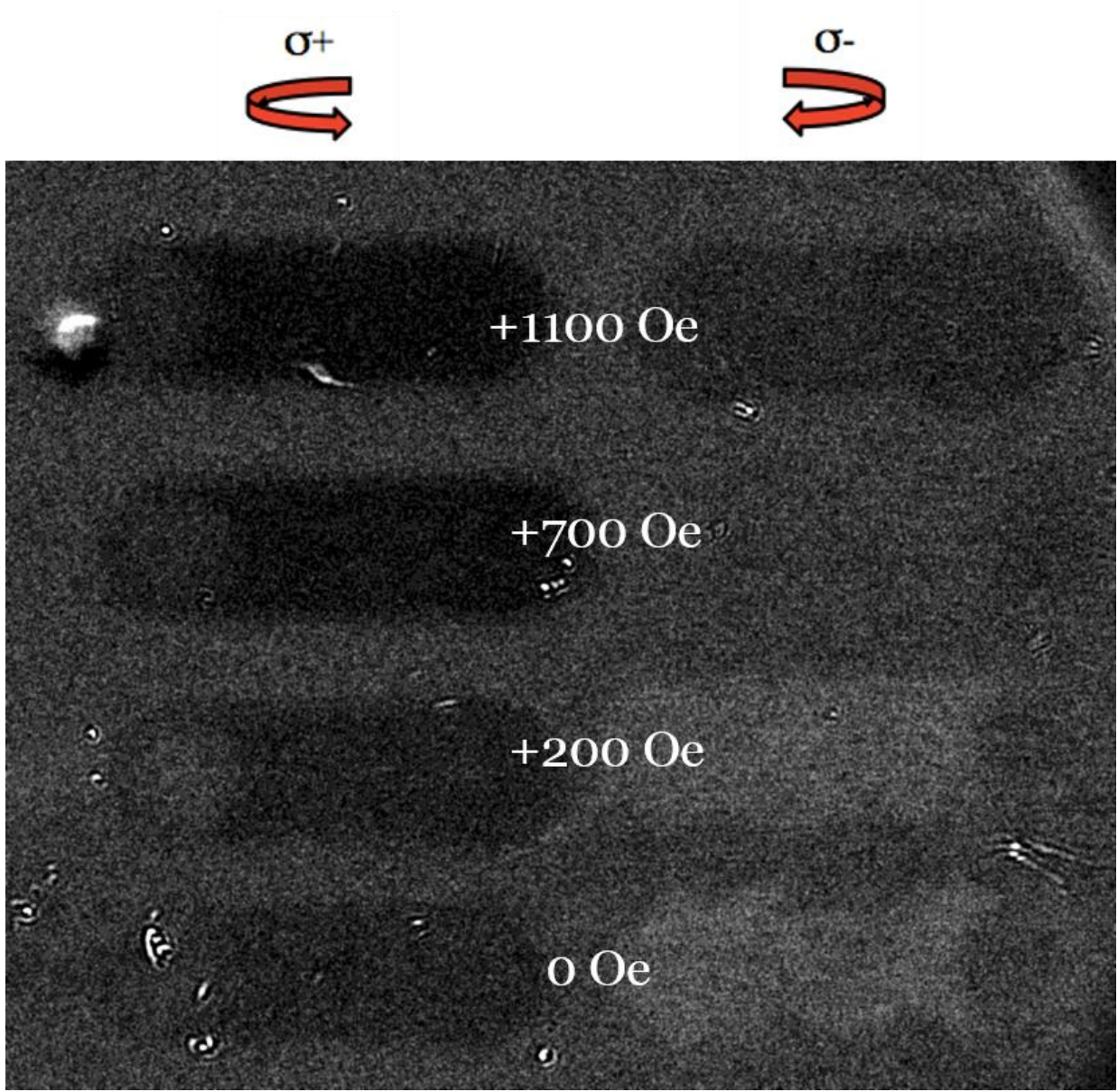

**Figure 4:** Magneto-photonic response in various applied magnetic field of a 15-nm FePtAgC granular film sample starting with an initially demagnetized sample. Shown are line scans for right circularly polarized light (σ+) in the left column and left circularly polarized light (σ-) in the right column. The laser power was 677 nW. The magnitude of the magnetic field is given in the figures and the orientation of the field supports the right circular polarization and opposes the left circular polarization.

# Supplementary Information

We further explored the effective driving energy for laser-induced AO-HDS and TD by applying an external magnetic fields during the experiments on the samples shown in Figs. 1a-c. In Supplementary Fig. S1 are the results obtained on the [Co(0.4 nm)/Pt(0.7nm)]$_8$ multilayer sample presented in Fig 1a of the main manuscript. At room temperature this sample exhibits a square hysteresis loop with a coercive field of 368 Oe. What is shown are line scans for different helicities (left and right circular polarization and linear polarization) with increasing applied magnetic fields. In zero applied field the sample shows TD as seen in Fig. 1a but with increasing applied field the contrast in the area where the laser is swept increases. A field on the order of 35-40 Oe is needed to saturate the film. When comparing left circular polarized light with linear polarized light that have same applied magnetic fields the results are very similar indicating the helicity of the light is not contributing to the final state of the magnetization.

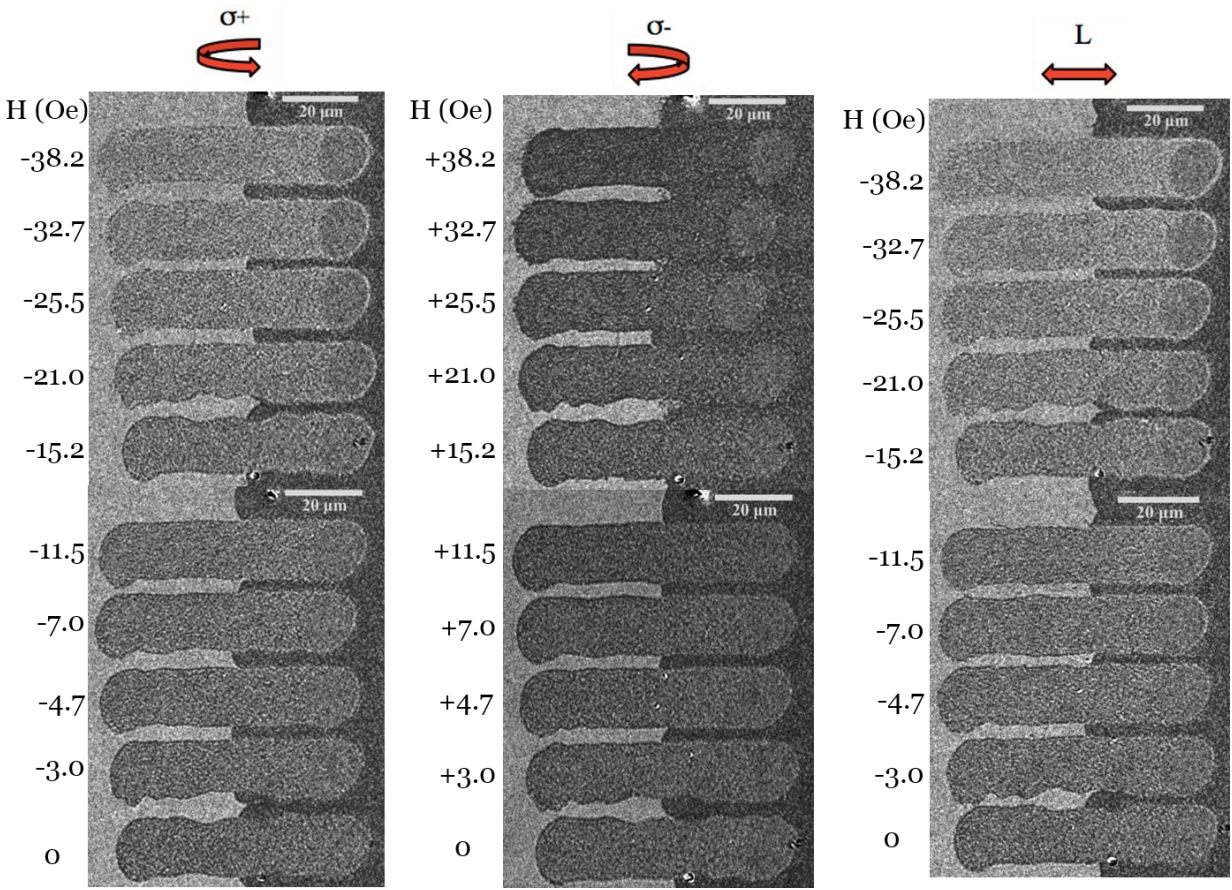

**Supplementary Figure S1:** Magnetic response of a [Co(0.4 nm)/Pt(0.7nm)]$_8$ multilayer to the combined effect of optical excitation and an external magnetic field starting with two domains and a domain well in the middle of each image. The magnitude of the magnetic field is shown to the left of each line scan and the average laser power was 682 nW.

In Supplementary Fig. S2 are the results for the [Co(0.4 nm)/Pt(0.7nm)]$_5$ multilayer sample shown in Fig 1b of the main manuscript. At room temperature this sample exhibits a square hysteresis loop with a coercive field of 297 Oe. Line scans for different helicities (left and right circular polarization and linear polarization) with increasing magnetic fields are shown. In the left image (σ+) the helicity of the light and the magnetic support the magnetization of the sample. Conversely for the middle images (σ-) the applied field opposes the helicity of the light. The right images are for linear polarization. When the helicity of the light supports the external magnetic field (σ+) an external magnetic field of 7 Oe is needed to saturate the film. Conversely for same applied field and the opposite helicity (σ-) requires an external field of 21-25 Oe to achieve saturation indicating a clear helicity dependent response. For linear light an external field of ~ 12 Oe is needed to achieve saturation.

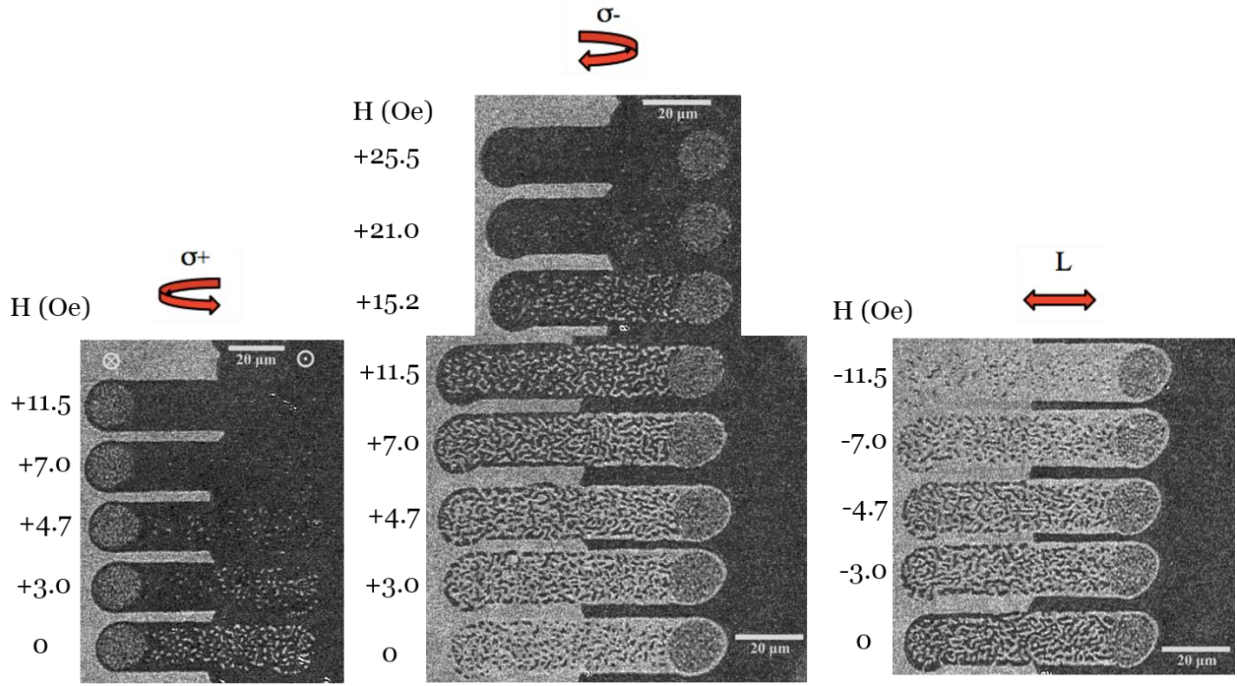

**Supplementary Figure S2:** Magnetic response of a [Co(0.4 nm)/Pt(0.7)]$_5$ multilayer to the combined effect of optical excitation and an external magnetic field starting with two domains and a domain well in the middle of each image. The magnitude of the magnetic field is shown to the left of each line scan and the average laser power was 720 nW.

In Supplementary Fig. S3 are the results for the [Co(0.4 nm)/Pt(0.7nm)]$_3$ multilayer sample shown in Figs. 1c and 1d of the main manuscript. At room temperature this sample exhibits a square hysteresis loop with a coercive field of 204 Oe. Line scans for different helicities (left and right circular polarization and linear polarization) are shown for increasing magnetic fields. In the left image (σ+) and the middle image (σ-) the applied field opposes the helicity of the light. When the helicity of the external magnetic field opposes the helicity of the light an external magnetic field of 7 Oe is yields a demagnetized film while 11.5 Oe will saturate the film in the field direction and opposite to the AO-HDS direction. For linear light an external field of ~ 3 Oe is needed to achieve saturation.

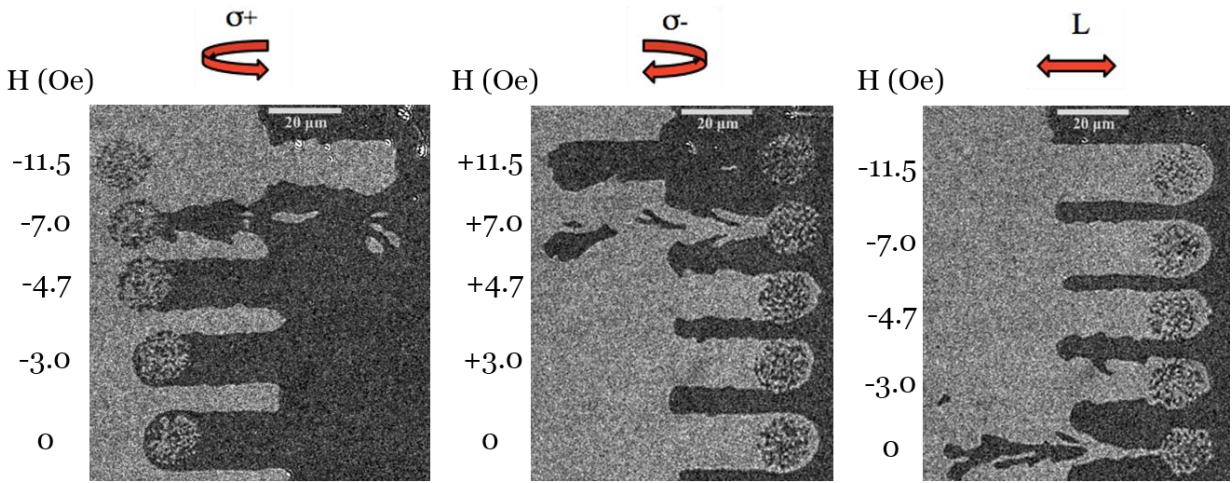

**Supplementary Figure S3:** Magnetic response of a [Co(0.4 nm)/Pt(0.7nm)]$_3$ multilayer to the combined effect of optical excitation and an external magnetic field starting with two domains and a domain well in the middle of each image. The magnitude of the magnetic field is shown to the left of each line scan and the average laser power was 523 nW.

Shown in Supplementary Figs. S4 and S5 are the electron microscopy and magnetometry results for the FePt films studied.

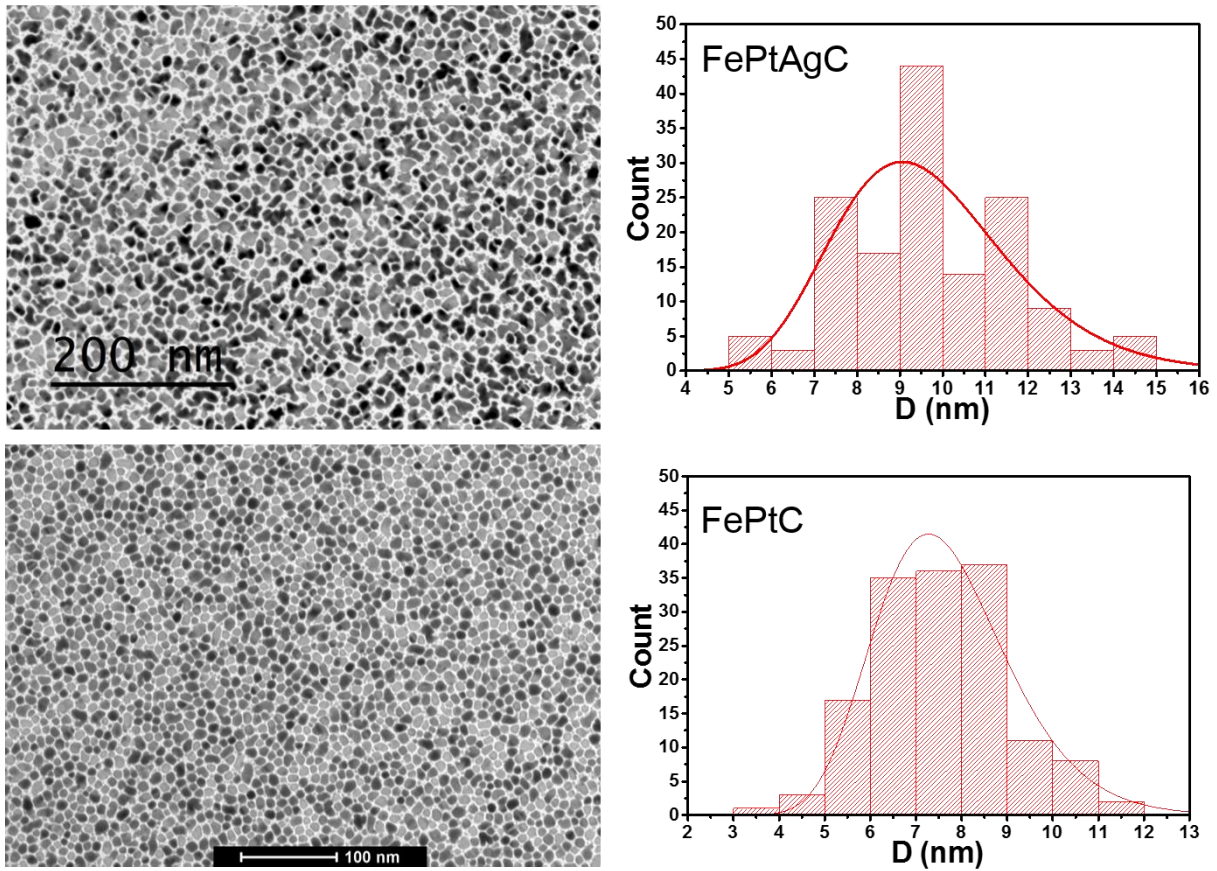

**Supplementary Figure S4:** Transmission electron microscopy (TEM) plan-view images (left) and analysis (right) of the FePtAgC granular film (upper) and FePtC granular film (lower). The right images show the analysis of the grain diameter distribution determined from each TEM image. For the FePtAgC film the mean grain diameter is 9.7 $\pm$ 2.1 nm and the average pitch distance is 15.5 $\pm$ 2.9 nm. For the FePtC film the mean grain diameter is 7.7 $\pm$ 2.1 nm and the average pitch distance is 10.8 $\pm$ 1.8 nm.

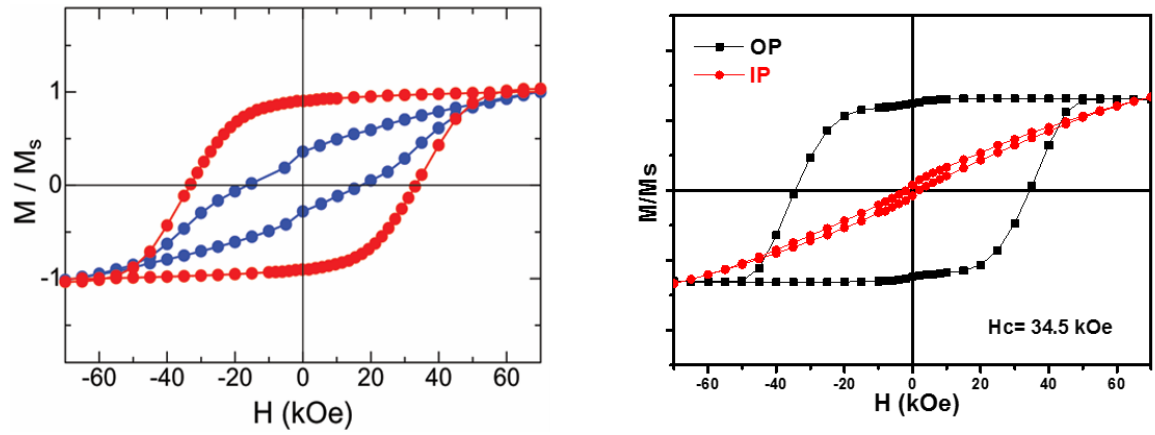

**Supplementary Figure S5:** Room-temperature magnetometry measurements for the FePtAgC (left) and FePtC (right) samples showing perpendicular magnetic anisotropy with coercive fields of ~3.5 T.

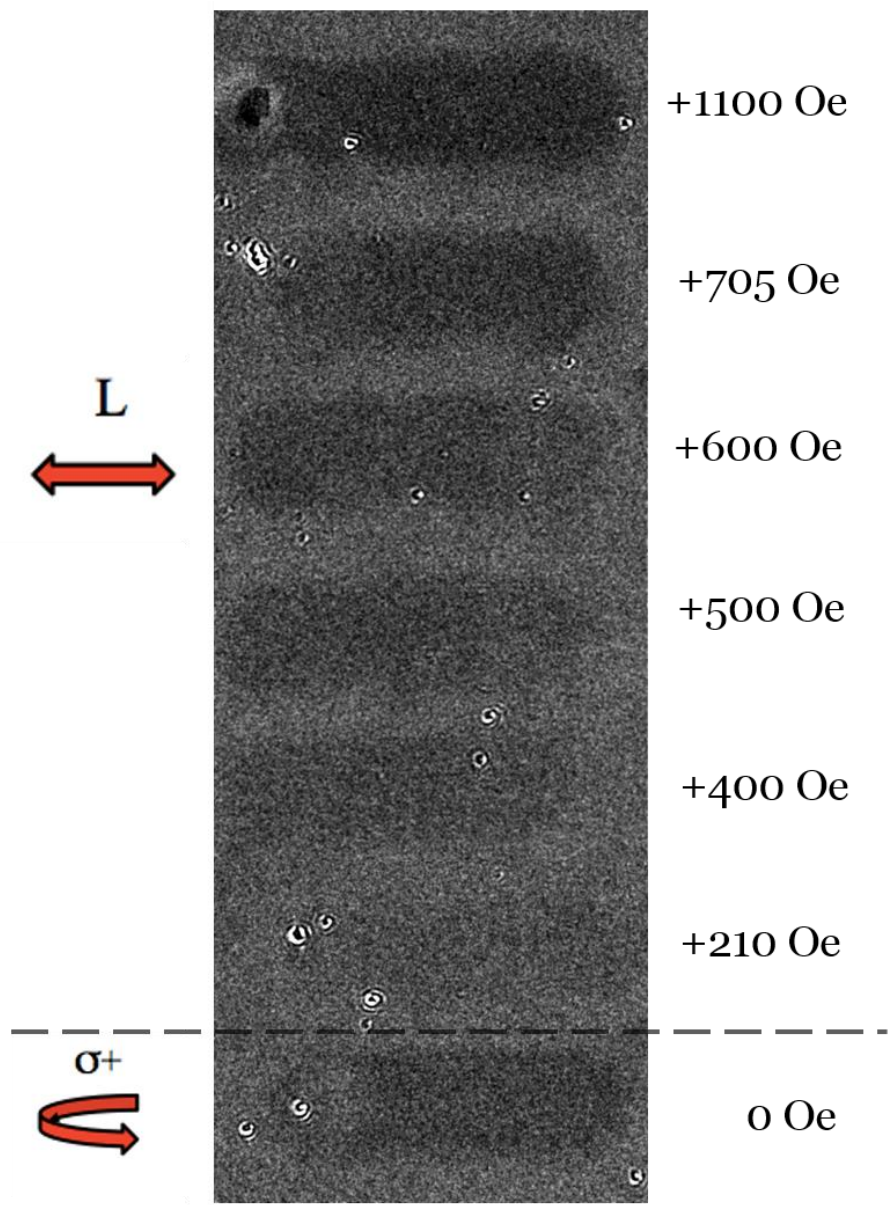

**Supplementary Figure S6:** Magneto-optical response of the FePtAgC sample. Shown are a line scan for right circular polarization in zero applied field (bottom) and line scans for linear polarization with increasing applied magnetic field. The magnitude of the field is shown in the right of the image.

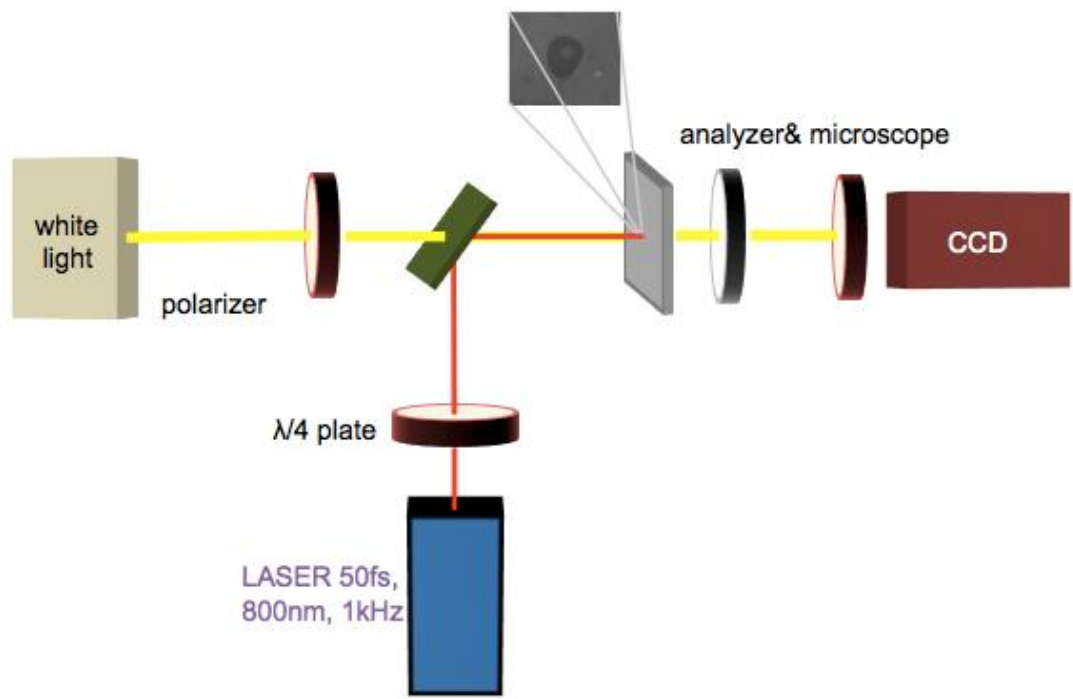

**Supplementary Figure S7:** Schematic of magnetic measurements apparatus showing a 50fs laser exiting the sample and the domain structure imaged using a Faraday microscope.